\documentclass[a4paper]{article}

\usepackage{amsmath}
\usepackage{amssymb}
\usepackage{amsthm}
\usepackage{color}
\usepackage{graphicx}
\usepackage{tikz}
\usepackage{xspace}
\usepackage{multirow}
\usepackage{amsmath, amsthm}
\usepackage{a4wide}
\usepackage{aliascnt}	%because autoref sees lemmas as theorems

\usetikzlibrary{shapes,snakes}

\usepackage[color=green!40]{todonotes}
\usepackage[pdfdisplaydoctitle,menucolor=orange!40!black,filecolor=magenta!40!black,urlcolor=blue!40!black,linkcolor=red!40!black,citecolor=green!40!black,colorlinks]{hyperref}

\usepackage[noend]{algpseudocode}
\usepackage{algorithm}

\makeatletter
\def\NAT@spacechar{~}% NEW
\makeatother

\newcommand{\maxopinfluence}{{\sc{Max Open $k$-Influence}}\xspace}
\newcommand{\maxclinfluence}{{\sc{Max Closed $k$-Influence}}\xspace}
\newcommand{\minclinfluence}{{\sc{Min Closed $k$-Influence}}\xspace}
\newcommand{\minopinfluence}{{\sc{Min Open $k$-Influence}}\xspace}
\newcommand{\decminopinfluence}{{\sc{Open $k$-Influence$_{\leq}$}}\xspace}
\newcommand{\decminclinfluence}{{\sc{Closed $k$-Influence$_{\leq}$}}\xspace}

\newcommand{\mcs}{{\sc{Monotone Circuit Satisfiability}}\xspace}

\newcommand{\clique}{{\sc{Clique}}\xspace}
\newcommand{\stable}{{\sc{Independent Set}}\xspace}

\newcommand{\TSS}{\textsc{Target Set Selection}\xspace}

\newlength{\atextwidth}
\newcommand{\N}{\mathbb{N}}
\newcommand{\R}{\mathbb{R}}
\DeclareMathOperator{\thr}{thr}

\newcommand{\act}[1]{\ensuremath{\mathcal{A}_{#1}}}
\newcommand{\rnd}[1]{\ensuremath{r_{#1}}}
\def \eps {\varepsilon}
\newcommand{\p}{\mathsf{P}}
\newcommand{\np}{\mathsf{NP}}

\newcommand{\fpt}{\mathsf{FPT}}
\newcommand{\wone}{\ensuremath{\mathsf{W[1]}}}
\newcommand{\wtwo}{\mathsf{W[2]}}

\newcommand{\wx}[1]{\mathsf{W[#1]}}

{\rm\bfseries}{\itshape}

\newcommand{\mytabref}[1]{[\hyperref[#1]{Th.~\ref*{#1}}]}

\newcommand{\ie}{\textit{i.e.}~}

\tikzstyle{vertex}=[circle, fill=white, draw, inner sep=2pt, minimum width=4pt]
\tikzstyle{edge} = [draw,-]

\pgfdeclarelayer{bg}
\pgfsetlayers{bg,main}

\newcommand{\problemopt}[3]{
  \vspace{1mm}
\noindent\fbox{
  \begin{minipage}{\atextwidth}
  #1 \\
  {\bf{Input:}} #2  \\
  {\bf{Output:}} #3
  \end{minipage}
  }
  \vspace{1mm}
}

\usepackage{etoolbox}

\newcounter{Bew1}
\newcounter{Bew2}
\newcounter{Def1}

\newtheorem{theorem}{Theorem}
\newtheorem{definition}{Definition}
\newaliascnt{lemma}{theorem}		%for autoref
\newtheorem{lemma}[theorem]{Lemma}
\aliascntresetthe{lemma}		%for autoref
\newtheorem{corollary}[theorem]{Corollary}
\newtheorem{proposition}[theorem]{Proposition}

\begin{document}
%\maketitle

%\rhauthor{Bazgan, Chopin, Nichterlein, Sikora}
%\rhtitle{Parameterized Inapproximability of Target Set Selection and Generalizations}

\title{Parameterized Inapproximability of Target Set Selection and Generalizations
\footnote{An extended abstract was accepted for publication in the Proceedings of the 10th conference Computability in Europe (CiE 2014)}}

%\author {Cristina Bazgan\\
%\affiliation{PSL, Universit\'{e} Paris-Dauphine, LAMSADE UMR CNRS 7243, France\\
%Institut Universitaire de France\\
%{\tt bazgan@lamsade.dauphine.fr}\\
%}
%\and Morgan Chopin\\
%\affiliation{Institut f\"ur Optimierung und Operations Research, Universit\"at Ulm, Germany\\
%{\tt morgan.chopin@uni-ulm.de}\\
%}
%\and Andr\'e Nichterlein\\
%\affiliation{Institut f\"ur Softwaretechnik und Theoretische Informatik, TU Berlin, Germany\\
%{\tt andre.nichterlein@tu-berlin.de}\\
%}
%\and Florian Sikora\\
%\affiliation{PSL, Universit\'{e} Paris-Dauphine, LAMSADE UMR CNRS 7243, France\\
%{\tt florian.sikora@lamsade.dauphine.fr}\\
%}
%}

\author{Cristina Bazgan\footnote{LAMSADE - CNRS UMR 7243, PSL, Universit\'e Paris-Dauphine (France). \texttt{\{bazgan,chopin,florian.sikora\}@lamsade.dauphine.fr}}~~\footnote{\noindent Institut Universitaire de France}
\and Morgan Chopin\footnote{\noindent Institut f\"ur Optimierung und Operations Research, Universit\"at Ulm, Germany. \texttt{morgan.chopin@uni-ulm.de}}
\and Andr\'e Nichterlein\footnote{\noindent Institut f\"ur Softwaretechnik und Theoretische Informatik, TU Berlin, Germany. \texttt{andre.nichterlein@tu-berlin.de}}
\and Florian Sikora$^{\dagger}$
}

\date{}

\maketitle

\begin{abstract}
% In this paper, we consider the problem of maximizing the spread of influence through a social network according to a given propagation rule.
In this paper, we consider the \TSS problem: given a graph and a threshold value $\thr(v)$ for each vertex $v$ of the graph, find a minimum size vertex-subset to ``activate'' such that all vertices of the graph are activated at the end of the propagation process.
A vertex $v$ is activated during the propagation process if at least $\thr(v)$ of its neighbors are activated.
This problem models several practical issues like faults in distributed networks or word-to-mouth recommendations in social networks.
% We show that this problem  is not fpt cost $g(k)$-approximable with respect to the parameter $k$, for any function $g$, unless $\fpt = \wx{P}$, even for restricted thresholds (namely constant and majority thresholds).
We show that for any functions~$f$ and~$\rho$ this problem cannot be approximated within a factor of~$\rho(k)$ in~$f(k) \cdot n^{O(1)}$ time, unless $\fpt = \wx{P}$, even for restricted thresholds (namely constant and majority thresholds), where $k$ is the number of vertices to activate in the beginning.
We also study the cardinality constraint maximization and minimization versions of the problem for which we prove similar hardness results.
%\todo{make a longer abstract?}
\end{abstract}

\section{Introduction}
Diffusion processes in graphs have been intensively studied~\cite{peleg02,kempe2003,chen2009,aazami2009,dreyer2009,chang2009,benzwi2011,reddy2011}.
One model to represent them is to define a \textit{propagation rule} and choose a subset of vertices that, according to the given rule, activates all or a fixed fraction of the vertices where initially all but the chosen vertices are inactive.
This models problems such as the spread of influence or information in social networks via word-of-mouth recommendations, of diseases in populations, or of faults in distributed computing~\cite{peleg02,kempe2003,dreyer2009}.
% to select an initial subset of vertices to activate (the other vertices are inactivate) in a graph such that, according to a \textit{propagation process}, the remaining vertices (or a fixed fraction of them) are activated.
% For example, if we consider an activated vertex as an infected person (by the \textit{flu}, \textit{smallpox} or \textit{tularemia}), then the goal for a bioterrorist is to select the minimum number of initial individuals to infect such that the propagation of the disease infect the largest number of persons.\todo{We should talk about social network since it is in our title}
One representative problem that appears in this context is the \textsc{Influence Maximization} problem introduced by Kempe~et al.~\cite{kempe2003}.
Given a directed graph and an integer $k$, the task is to choose a vertex subset of size at most $k$  such that the number of activated vertices at the end of the propagation process is maximized.
The authors show that the problem is polynomial-time %($\frac{e}{e-1} + \eps$)-approximable
($\frac{e}{e-1} + \eps$)-approximable for any $\eps > 0$ under some stochastic propagation rules, but $\np$-hard to approximate within a ratio of $n^{1-\eps}$ for any $\eps > 0$ for general propagation rules.

In this paper, we use the following deterministic propagation model.
We are given an undirected graph, a threshold value~$\thr(v)$ associated to each vertex $v$, and the following propagation rule: a vertex becomes active if at least~$\thr(v)$ many neighbors of~$v$ are active.
The propagation process proceeds in several rounds and stops when no further vertex becomes active.
Given this model, finding and activating a minimum-size vertex subset such that all the vertices become active is known as the \TSS problem and was introduced by Chen~\cite{chen2009}.

\TSS has been shown $\np$-hard even for bipartite graphs of bounded degree when all thresholds are at most two~\cite{chen2009}.
Moreover, the problem was shown to be hard to approximate  in polynomial time within a ratio $O(2^{\log^{1-\eps}n})$ for any $\eps > 0$, even for constant degree graphs with thresholds at most two and for general graphs when the threshold of each vertex is half its degree (called \textit{majority} thresholds) \cite{chen2009}.
If the threshold of each vertex equals its degree (\textit{unanimity} thresholds), then the problem is equivalent to the vertex cover problem~\cite{chen2009} and, thus, admits a $2$-approximation and is hard to approximate with a ratio better than $1.36$~\cite{DS02}.
Concerning the parameterized complexity, the problem is known to be~$\wtwo$-hard with respect to (w.r.t.) the solution size, even on bipartite graphs of diameter four with majority thresholds or thresholds at most two \cite{NNUW12}.
Furthermore, it is~$\wone$-hard w.r.t. each of the parameters ``treewidth'', ``cluster vertex deletion number'', and ``pathwidth'' \cite{benzwi2011,CNNW12}.
On the positive side, the problem becomes fixed-parameter tractable w.r.t. each of the single parameters ``vertex cover number'', ``feedback edge set size'', and ``bandwidth'' \cite{NNUW12,CNNW12}.
If the input graph is complete, has a bounded cliquewidth, or has a bounded treewidth and bounded thresholds then the problem is polynomial-time solvable~\cite{NNUW12,benzwi2011,Cicalese2013}.

% Since the problem is strongly inapproximable in polynomial time and also hard for the parameterized complexity point of view,
Motivated by the hardness of approximation and parameterized hardness we showed in previous work \cite{BCNS12} that the cardinality constraint maximization version of \TSS, that is to find a fixed number $k$ of vertices to activate such that the number of activated vertices at the end is maximum, is strongly inapproximable in fpt-time w.r.t. the parameter $k$, even for restricted thresholds.
For the special case of unanimity thresholds, we showed that the problem is still inapproximable in polynomial time, but becomes $r(n)$-approximable in fpt-time w.r.t. the parameter $k$, for any strictly increasing function~$r$.

Continuing this line of research, we study in this paper \TSS and its variants where the parameter relates to the optimum value.
This requires the special definition of ``fpt cost approximation'' since in parameterized problems the parameter is given which is not the case in optimization problems (see \autoref{sec:preliminaries} for definitions).
Fpt approximation algorithms were introduced in \cite{chen2006,downey2006,CH2010}, see also the survey of Marx \cite{marx06}.
Besides this technical difference observe that \TSS can be seen as a special case of the previously considered problem, since activating all vertices is a special case of activating a given number of vertices.
Strengthening the known inapproximability results, we first prove in \autoref{sec:tss} that \TSS is not fpt cost $\rho$-approximable, for any computable function $\rho$, unless $\fpt = \wx{P}$,
even for majority and constant thresholds.
% First, we strengthen in \autoref{sec:tss} the inapproximability of \TSS by showing that the problem is not fpt cost $\rho$-approximable, for any computable function $\rho$, unless $\fpt = \wx{P}$.
Complementing our previous work, we also study in \autoref{sec:maxminkinfluence} the cardinality constraint maximization and minimization versions of \TSS.
We prove that these two problems are not fpt cost $\rho$-approximable, for any computable function $\rho$,  unless $\fpt = \wone$.
Note that we study the parameterized approximability of these problems in a different way as in our previous work~\cite{BCNS12}. Indeed, here, the parameter is related to the solution size (i.e. the number of vertices activated in the end), while in our previous work, the parameter was the size of the set of vertices initially activated.

\section{Preliminaries and basic observations} \label{sec:preliminaries}

In this section, we provide basic backgrounds and notation used throughout this paper and define \TSS.
For details on parameterized complexity we refer to the monographs \cite{DF99,FG06,Nie06}. For details on parameterized approximability we refer to the survey of Marx \cite{marx06}.

\paragraph{Graph terminology.} Let $G=(V,E)$ be an \textit{undirected graph}.  For a subset $S\subseteq V$, $G[S]$ is the subgraph induced by $S$. The \textit{open neighborhood} of a vertex $v \in V$ in $G$, denoted by $N_G(v)$, is the set of all neighbors of $v$ in $G$. The \textit{closed neighborhood} of a vertex $v$ in $G$, denoted $N_G[v]$, is the set $N_G(v) \cup \{v\}$.
%Furthermore, for a vertex set~$V' \subset V$ we set~$N_G(V') = \bigcup_{v\in V'} N_G(v)$\todo{MC:better $N(V') = N[V'] \setminus S$? here $N(V')$ may include $V'$ e.g. $V'$ is a clique.} and~$N_G[V'] = \bigcup_{v\in V'} N_G[v]$.
%\todo[inline]{Cristina : do we need this definition ?}
%The set $N^{k}[v]$, called the $k$-neighborhood of $v$, denotes the set of vertices which are at distance at most $k$ from $v$ (thus $N^1[v]= N[v]$).
The \textit{degree} of a vertex $v$ is denoted by $\deg_G(v)$ and the \emph{maximum degree} of the graph~$G$ is denoted by~$\Delta_G$.
We skip the subscripts if~$G$ is clear from the context.
%Two vertices are \textit{twins} if they have the same neighborhood. They are called \textit{true twins} if they are moreover neighbors, \textit{false twins} otherwise.

%\paragraph{Cardinality constrained problem.} The problems studied in this paper are cardinality constrained. We use the notations and definitions from Cai~\cite{cai08}. A cardinality constrained
%optimization problem is a quadruple $A=(\mathcal{B}, \Phi, k, obj)$, where $\mathcal{B}$ is a finite set called solution base, $\Phi: 2^{\mathcal{B}} \to \{0,1,2,\ldots\} \cup \{-\infty,+\infty\}$
%an objective function, $k$ a non-negative integer and $obj \in \{min, max\}$. The goal is then to find a solution $S \subseteq \mathcal{B}$ of cardinality $k$ so as to maximize (or minimize)
%the objective value $\Phi(S)$. If $S$ is not a feasible solution we set $\Phi(S) = -\infty$ if $obj = max$ and $\Phi(S) = +\infty$ otherwise.

\paragraph{Parameterized complexity.}
\sloppy A parameterized problem $(I,k)$ is said \textit{fixed-parameter tractable} (or in the class $\fpt$) w.r.t.
parameter $k$ if it can be solved in $f(k)\cdot|I|^c$ time, where $f$ is any computable function and $c$ is a constant.
% \todo{Cristina: the previous definition was not correct since $n$ was not define. I prefer this, that was the old one. - A.: agree}
%A problem is said \textit{fixed-parameter tractable} (or in the class $\fpt$) w.r.t.
% a parameter~$k$ if it can be solved exactly in~$f(k)\cdot n^{O(1)}$ time, where $f$ is any computable function.
% Parameterized algorithms do not necessarily exist, it is possible to prove that a problem do not have a parameterized algorithm via fpt-reductions. The parameterized complexity hierarchy is composed of the classes $\fpt \subseteq \wone \subseteq \wtwo \subseteq \dots \subseteq \mathsf{W[P]}$. It is unlikely to find parameterized algorithm for a problem that is hard in $\wone$ under fpt-reduction.
The parameterized complexity hierarchy is composed of the classes $\fpt \subseteq \wone \subseteq \wtwo \subseteq \dots \subseteq \mathsf{W[P]}$.
A $\wone$-hard problem is not fixed-parameter tractable (unless $\fpt=\wone$) and one can prove  the $\wone$-hardness by means of a \emph{parameterized reduction} from a $\wone$-hard problem.
Such a reduction between two parameterized problems $A_1$ and $A_2$ is a mapping of any instance~$(I,k)$ of~$A_1$ in $g(k)\cdot |I|^{O(1)}$ time
(for some computable function~$g$) into an instance $(I',k')$ for~$A_2$ such that $(I,k)\in A_1\Leftrightarrow (I',k')\in A_2$ and $k'\le h(k)$
for some function~$h$.

%\todo{Approx. paragraph and def. 1 may be deleted in the conf. version}
\paragraph{Parameterized approximation.}  An $\np$-optimization problem $Q$ is a \sloppy tuple $({\cal I}, Sol, val, goal)$, where ${\cal I}$ is the set of instances, $Sol(I)$ is the set of feasible solutions for instance $I$, $val(I,S)$ is the value of a feasible solution $S$ of $I$, and $goal$ is either max or min.
We assume that $val(I,S)$ is computable in polynomial time and that $|S|$ is polynomially bounded by $|I|$ \ie $|S| \leq |I|^{O(1)}$.

\begin{definition}[fpt cost $\rho$-approximation algorithm, Chen et al.~\cite{chen2006}]
Let $Q$ be an optimization problem  and $\rho\colon \N \rightarrow \R$ be a function such that $\rho(k) \geq 1$ for every $k\geq 1$ and  $k \cdot \rho(k)$  is nondecreasing (when $goal$ = min)
or %$k/\rho(k)$
$\frac{k}{\rho(k)}$
 is unbounded and nondecreasing  (when $goal$ = max).
%The following definition was introduced by Chen et al.~\cite{chen2006}.
 A decision algorithm ${\cal A}$ is an \emph{fpt cost $\rho$-approximation algorithm} %\todo{marx call this fpt COST apprx}
 for $Q$ (when $\rho$ satisfies the previous conditions) if for every instance $I$ of $Q$ and integer $k$, with $Sol(I)\neq \emptyset$, its output satisfies the following conditions:

\begin{enumerate}
	\item If $opt(I) > k$ (when $goal$ = min) or $opt(I) < k$ (when $goal$ = max), then ${\cal A}$ rejects $(I,k)$.
	\item If $k\geq opt(I) \cdot \rho(opt(I))$ (when $goal$ = min) or %$k\leq opt(I) /\rho(opt(I))$
	$k\leq \frac{opt(I)}{\rho(opt(I))}$
	 (when $goal$ = max), then ${\cal A}$ accepts $(I,k)$.
\end{enumerate}
 %\vspace{-0,2cm}	%beurk vspace
 Moreover the running time of ${\cal A}$ on input $(I,k)$ is $f(k) \cdot |I|^{O(1)}$.
%\todo{Should we keep $\rho(opt)$ in the definition or put $\rho(|I|)$ instead ?}
%
If such  a decision algorithm ${\cal A}$ exists then $Q$ is called fpt cost $\rho$-approximable.
\end{definition}

% \todo[inline]{Florian: I would add a word that this definition is weaker that the original one,
% but since we show only hardness results, our results are stronger...Cristina : I am not sure that we have to introduce also this notion. A: I would also not define the other notion if we don't use it...}

% \medskip
% \appendixDefinition
{
The notion of a gap-reduction was introduced in \cite{AL96} by Arora and Lund.
We use in this paper a variant of this notion, called fpt gap-reduction.

\begin{definition}[fpt gap-reduction]
A problem~$A$ parameterized by~$k$ is called {\it fpt gap-reducible} to an optimization problem $Q$ with gap~$\rho$ if
for any instance $(I,k)$ of $A$ we can construct an instance $I'$ of $Q$
in~$f(k)\cdot|I|^{O(1)}$ time while satisfying the following properties:
\begin{enumerate}
\item If~$I$ is a yes instance then $opt(I') \leq \frac{g(k)}{\rho(opt(I'))}$ (when $goal$ = min) or $opt(I') \geq g(k) \rho(opt(I'))$ (when $goal$ = max),
\item If $I$ is a no instance then $opt(I') > g(k)$ (when $goal$ = min) or $opt(I') < g(k)$ (when $goal$ = max),
\end{enumerate}
for some function~$g$. The function~$\rho$ satisfies the aforementioned conditions.
\end{definition}

The interest of the fpt gap-reduction is the following result that immediately follows from the previous definition:

\begin{lemma} \label{lem:fpt-gap}
If a parameterized problem~$A$ is~$\mathcal{C}$-hard and fpt gap-reducible to an optimization problem $Q$ with gap~$\rho$ then~$Q$ is not fpt cost $\rho$-approximable unless~$\fpt = \mathcal{C}$ where $\mathcal{C}$ is any class of the parameterized complexity hierarchy.
\end{lemma}
}

\paragraph{Problem statement.}
Let~$G=(V,E)$ be an undirected graph and let $\thr\colon V\rightarrow\N$ be a threshold
function such that $1\leq \thr(v) \leq \deg(v)$, for all $v\in V$.
The definition of \TSS is based on the notion of
``activation''.  Let~$S\subseteq V$.  Informally speaking, a
vertex~$v\in V$ gets activated by~$S$ in the $i^\text{th}$ round if at
least~$\thr(v)$ of its neighbors are active after the previous round
(where~$S$ are the vertices active in the $0^\text{th}$ round).  Formally,
for a vertex set~$S$, let~$\act{G,\thr}^i(S)$ denote the set of vertices
of~$G$ that are \emph{activated by~$S$ at the~$i^\text{th}$ round}, with
\begin{align*}
  \act{G,\thr}^0(S)    &=  S \text{ and } \\
  \act{G,\thr}^{i+1}(S) &=  \act{G,\thr}^i(S)\cup\{v\in V\colon |N(v)\cap\act{G,\thr}^i(S)|\geq\thr(v)\}.
\end{align*}
% $\act{G,\thr}^0(S) =  S$ and~$\act{G,\thr}^{i+1}(S) = \act{G,\thr}^i(S)\cup\{v\in V : |N(v)\cap\act{G,\thr}^i(S)|\geq\thr(v)\}.$
For~$S\subseteq V$, the unique positive integer~$r$ with~$\act{G,\thr}^{r-1}(S)\not=\act{G,\thr}^r(S) = \act{G,\thr}^{r+1}(S)$ is called the \emph{number~$\rnd{G}(S)$ of activation rounds}.
It is easy to see that~$\rnd{G}(S)\leq |V(G)|$ for all graphs~$G$.
Furthermore, we call~$\act{G,\thr}(S)=\act{G,\thr}^{\rnd{G}(S)}(S)$ the set of vertices that are \emph{activated by~$S$}.
If~$\act{G,\thr}(S)=V$, then~$S$ is called a \emph{target set for~$G$}.
%\TSS is the problem of finding  a minimum-cardinality target set.
\TSS is formally defined as follows.

\problemopt
{\TSS}
{A graph $G=(V,E)$ and a threshold function $\thr\colon V \to \N$.}
{A target set for~$G$ of minimum cardinality.}

%Since we also study variants of this problem, we state the decision version of the most general problem variant.
We also consider the following cardinality constrained version.

%\problemopt{\maxclinfluence}{A graph $G=(V,E)$, a threshold function $\thr:V \to \N$, and integers $k$ and~$\ell$.}{A subset $S \subseteq V$, $|S| \leq k$ such that $|\act{G,\thr}(S)| \ge \ell$.}

%\problemdec{\maxclinfluence}{A graph $G=(V,E)$, a threshold function $\thr:V \to \N$, and integers $k$ and~$\ell$.}{Is there a  subset $S \subseteq V$, $|S| \leq k$ such that $|\act{G,\thr}(S)| \ge \ell$?}

\problemopt{\maxclinfluence}
{A graph $G=(V,E)$, a threshold function $\thr\colon V \to \N$, and an integer $k$.}
{A subset $S \subseteq V$ with~$|S| \leq k$ maximizing $|\act{G,\thr}(S)|$.}
% {A subset $S \subseteq V$ with~$|S| \leq k$ such that $|\act{G,\thr}(S)|$ is maximum.}

%If~$\ell = |V|$, then we have the \TSS problem as introduced by \citet{chen2009}.
The \maxopinfluence problem asks for a set $S\subseteq V$ with $|S| \leq k$  such that $|\act{G,\thr}(S) \setminus S|$ is maximum.
We remark that this difference in the definition is important when considering the approximability of these problems.
Finally, \minclinfluence (resp. \minopinfluence) is also defined similarly,
but one ask for a solution $S \subseteq V$ with $|S|= k$ such that $|\act{G,\thr}(S)|$ is minimum (resp.  $|\act{G,\thr}(S) \setminus S|$  is minimum).

%\todo[inline]{I changed the definition in a decision version and keep the previous one in comments in case you do not agree.
%
%-- A: agree.
%
% / I still think we should say that we define these problem as decision (which is a weeker version of the fpt approximation) because we %got only hardness results.
%
%-- A: I have no strong opinion here. We mentioned that we only have hardness in the intro but we can remind the reader here...}

% \problemdec{\TSS (TSS)}{An undirected graph~$G=(V,E)$, a threshold function $\thr\colon V\rightarrow\N$ and an integer~$k$.}{Is there a target set~$S\subseteq V$ for~$G$ that contains at most~$k$ vertices?}
%\todo{Cristina : The same remark, remove 0? A.N: Agree.}
%\todo{Should we keep parameter k ? If yes, should we put it also for \mcs ?}

%\todo{should we say that we opt $k$ for tss but $l$ for the 2 others?}

%\todo{why do we define tss here but not the influence ones?}

\paragraph{Directed edge gadget.} We will use the \emph{directed edge gadget} as used and proved by Chen~\cite{chen2009} throughout our work:
A directed edge gadget from a vertex $u$ to another vertex $v$ consists of a $4$-cycle $\{a,b,c,d\}$ such that $a$ and $u$ as well as~$c$ and $v$ are adjacent. Moreover $\thr(a) = \thr(b) = \thr(d) = 1$ and $\thr(c) = 2$ (see \autoref{fig:directed}).
% The idea is to ensure that the propagation goes in one direction \ie if $u$ is activated then so does $v$ but the converse is not possible.
The idea is that the vertices in the directed edge gadget become active if~$u$ is activated but not if~$v$ is activated.
Hence, the activation process may go from~$u$ to~$v$ via the gadget but not in the reverse direction. %\todo{currently a dirty copy\& paste solution. Introduce directed edge only once in the paper...}

%\appendixproof{Figure}
{
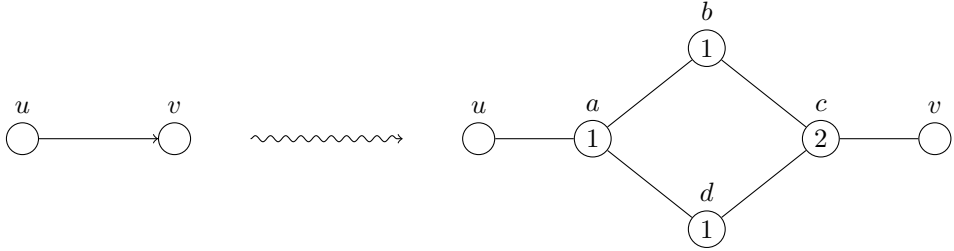
\begin{figure}[t]
\centering
\begin{tikzpicture}[scale=2,auto]
	\node[vertex,label=$u$,minimum width=12pt] (u) at (0, 0) {};%$\thr(u)$
	\node[vertex,label=$v$,minimum width=12pt] (v) at (1, 0) {};
	\draw[->] (u) -- (v);

	\draw[decorate, decoration={snake, amplitude=.4mm, segment length=2mm},->] (1.5,0) -- (2.5,0);
	\begin{scope}[xshift=3cm]
		\node[vertex,label=$u$,minimum width=12pt] (u) at (0, 0) {};%$\thr(u)$
		\node[vertex,label=$v$,minimum width=12pt] (v) at (3, 0) {};
		\node[vertex,label=$a$] (a) at (0.75, 0) {$1$};
		\node[vertex,label=$b$] (b) at (1.5, 0.6) {$1$};
		\node[vertex,label=$c$] (c) at (2.25, 0) {$2$};
		\node[vertex,label=$d$] (d) at (1.5, -.6) {$1$};
		\draw (u) -- (a) -- (b) -- (c) -- (d) -- (a);
		\draw (v) -- (c);
	\end{scope}
\end{tikzpicture}
\caption{Illustration of the directed edge gadget from $u$ to $v$.}\label{fig:directed}
\end{figure}
}

\section{Parameterized inapproximability of \TSS}\label{sec:tss}

Marx \cite{Mar13} showed that the \mcs problem admits no fpt cost~$\rho$-approximation algorithm for any function~$\rho$ unless~$\fpt=\wx{P}$.
%\todo{we should introduce the parameterized approximation notion of Marx since the parameter is what we want to approximate now  DONE}
In this section we show that we can transfer this strong inapproximability result from \mcs to \TSS.

Before defining \mcs, we recall the following notations.
A \emph{monotone (boolean) circuit} is a directed acyclic graph.
The nodes with in-degree at least two are labeled with \emph{and} or with \emph{or}, the~$n$ nodes with in-degree zero are input nodes, and due to the monotonicity there are no nodes with in-degree one (negation nodes in standard circuits).
Furthermore, there is one node with out-degree zero, called the \emph{output node}.
For an assignment of the input nodes with true/false, the circuit is satisfied if the output node is evaluated (in the natural way) to true.
The weight of an assignment is the number of input nodes assigned to true.
We denote an assignment as a set~$A \subseteq \{1, \ldots, n\}$ where $i \in A$ if and only if the~$i^{\text{th}}$ input node is assigned to true.
%
%
%The decision version associated to \mcs is defined as follows:
The \mcs problem is then defined as follows:
%

%\problemdec
\problemopt
	{\mcs}
%	{A monotone circuit~$C$ and an integer~$k$.}
  {A monotone circuit~$C$.}
%	{Is there a satisfying assignment with weight at most~$k$, that is, a satisfying assignment with at most~$k$ input nodes set to true?}
 	{A satisfying assignment of minimum weight, that is, a satisfying assignment with a minimum number of input nodes set to true.}

By reducing \mcs to \TSS in polynomial time such that there is a ``one-to-one'' correspondence between the solutions, we show the inapproximability result transfers to \TSS.
First, we show one reduction working with general thresholds. Using further gadgets, we  describe how to achieve constant or majority thresholds in our constructed instance.

\subsection{General thresholds}
% \paragraph{General thresholds.}
%
% \paragraph{Construction.}
As mentioned above, we will reduce from \mcs, and thus derive the same inapproximability result for \TSS as for \mcs.
% Next, we use and \emph{and gadget} with fan-in~$\ell$ is a vertex~$v$ with $\ell$ ingoing directed edges pointing to~$v$, one outgoing directed edge, and a threshold~$\thr(v) = \ell$.
% Observe that~$v$ becomes active if and only if all~$\ell$ vertices from the ingoing edges are active.
% Hence, $v$ acts like an \emph{and gate}.
% Similarly, an \emph{or gadget} with fan-in~$\ell$ is a vertex~$v$ with $\ell$ ingoing directed edges pointing to~$v$, one outgoing directed edge, and a threshold~$\thr(v) = 1$.
% Hence, $v$ acts like an \emph{or gate}.
%Now we show our construction.

%We assume without loss of generality that when assigning all input nodes to true will evaluate all and-nodes and or-nodes to true.%
%\todo{Cristina: why this phrase ? we supposed that the circuit is monotone, so this is clear true. A.N: Agree.}

%From \autoref{lem:monCircuitReduction} we can derive the same inapproximability result for \TSS as for \mcs.

%\todo{can't we just prove directly Th 1 with using a lemma?}
%

\begin{theorem}\label{thm:tss_inapprox}
%	There is \todo{There are no or There is no? Previously, we said ``there are no nodes with...''} no fpt cost~$g(k)$-approximation algorithm  parameterized by $k$ for \TSS, for any computable function $g$, unless $\fpt=\wx{P}$.
	\TSS is not fpt cost~$\rho$-approximable, for any computable function $\rho$, unless $\fpt=\wx{P}$.
\end{theorem}

% \todo{Cristina: I added "for any computable function $\rho$ "  in all innaproximability results. Do you agree ? A: yes.}
%
%\begin{lemma}\label{lem:monCircuitReduction}
%	Let~$(C,k)$ be a \mcs instance. Then, we can construct in polynomial time an instance $(G = (V,E),\thr,k')$ of \TSS such that
%\begin{enumerate}
%		\item[(i)]  for every satisfying assignment~$A$ for~$C$ there exists a target set of size~$|A|$ for~$G$, and
%		\item[(ii)]  for every target set~$S$ for~$G$ there exists a satisfying assignment of size at most~$|S|$ for~$C$.
%	\end{enumerate}
%\end{lemma}
% \appendixproof{\autoref{thm:tss_inapprox}}
{
\begin{proof}
Let~$C$ be an instance of \mcs. We construct an instance of \TSS as follows. Initialize~$G = (V,E)$ as a copy of the directed acyclic graph~$C$ where each directed edge is replaced by a directed edge gadget.
We call a vertex in~$G$ an input vertex (resp. output vertex, and-vertex, or-vertex) if it corresponds to an input node (resp. output node, and-node, or-node).
Next, for each and-node in~$C$ with in-degree~$d$
% \todo{maybe $\ell$ is not the best since it's used for something else? A: agree}
set the threshold of the corresponding and-vertex in~$G$ to~$d$ and for each or-vertex in~$G$ set the threshold to~$1$.
Set the threshold of each input vertex in~$G$ to~$n+1$.
Next, add~$n$ copies %\todo{Morgan: Maybe adding $n-1$ copies is sufficient since $|S|< n$?} of~$G$
to~$G$ and ``merge'' all vertices corresponding to the same input node. %, that is, after that step $G$ contains one vertex for each input node of~$C$ and~$n+1$ vertices each other node.
This means, that for an input node~$v$ with an outgoing edge~$(v,w)$ in~$C$ the graph~$G$ contains~$n+1$ vertices~$w_1, \ldots, w_{n+1}$ and $n+1$ directed edges from~$v$ to $w_i$, $1 \le i \le n+1$.
Finally, add directed edges from each output vertex to each input vertex.
% and set~$k'=k$ (not used).
This completes our construction (see~\autoref{fig:monCircuitReduction}).%\todo{I think all arrows represents directed edge gadget}
To complete the proof, it remains to show that

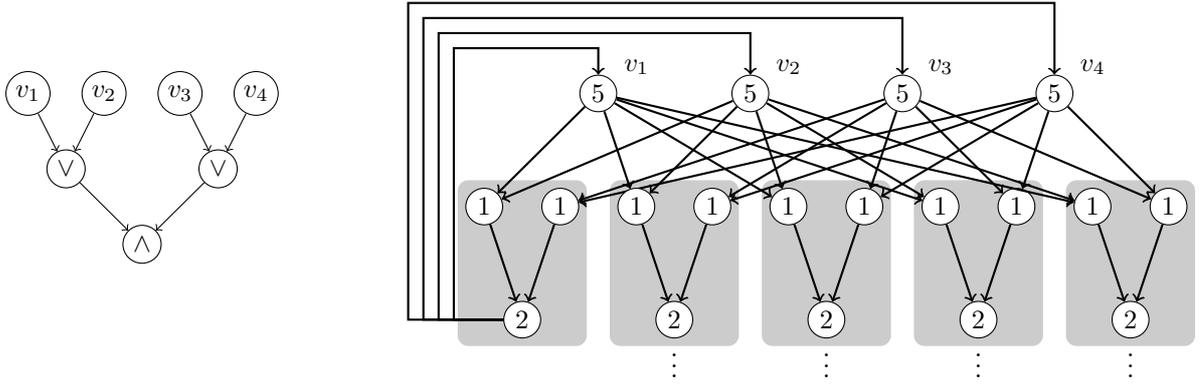
\begin{figure}[t]
\centering
%\resizebox{\textwidth}{!}{
	\begin{tikzpicture}[scale=1, auto,swap]
	%\tikzstyle{vertex}=[circle,draw,minimum size=20pt,inner sep=0pt,fill=white]

	\begin{scope}

	\node[vertex] (or1) at (1.5, -1) {$\vee$};
	\node[vertex] (or2) at (3.5, -1) {$\vee$};

	\node[vertex] (and1) at (2.5, -2) {$\wedge$};

	\foreach \i in {1,...,4} {
		\node[vertex] (v\i) at (\i, 0) {$v_{\i}$};
	}

	\path[edge,->] (v1) -- (or1);
	\path[edge,->] (v2) -- (or1);
	\path[edge,->] (v3) -- (or2);
	\path[edge,->] (v4) -- (or2);

	\path[edge,->] (or1) -- (and1);
	\path[edge,->] (or2) -- (and1);

	\end{scope}

	\begin{scope}[xshift=6cm]
	\foreach \i in {1,...,4} {
		\node[vertex,label=30:$v_{\i}$] (v\i) at (\i*2+0.5, 0) {$5$};
	}

	\foreach \i in {1,...,10} {
		\node[vertex] (or\i) at (\i, -1.5) {$1$};

			\ifthenelse{\isodd{\i}}{
			\path[edge,->,thick] (v1) -- (or\i);
			\path[edge,->,thick] (v2) -- (or\i);
		}{
			\path[edge,->,thick] (v3) -- (or\i);
			\path[edge,->,thick] (v4) -- (or\i);
		}
	}

	\foreach \i in {1,...,5} {
		\node[vertex] (and\i) at (\i*2-0.5, -3) {$2$};
			\ifthenelse{\i > 1}{
		\node (txt\i) at (\i*2-0.5, -3.5) {\vdots};
		}{}
		\pgfmathparse{int(\i*2-1)}  \let\a\pgfmathresult
		\pgfmathparse{int(\i*2)}  \let\b\pgfmathresult
		\path[edge,->,thick] (or\a) -- (and\i);
		\path[edge,->,thick] (or\b) -- (and\i);

			\begin{pgfonlayer}{bg}
				\path (and\i.south -| or\a.west)+(-0.1,-0.1) node (a-2) {};
			\path (or\b.north -| or\b.east)+(+0.1,+0.1) node (b-2) {};
			\path[rounded corners, fill=black!20]
					(a-2) rectangle (b-2);
		\end{pgfonlayer}
	}

	\draw [edge,->,thick] (and1) to (0.6,-3) to (0.6,0.6)  to (2.5,0.6)  to (v1);
	\draw [edge,->,thick] (and1) to (0.4,-3) to (0.4,0.8)  to (4.5,0.8)  to (v2);
	\draw [edge,->,thick] (and1) to (0.2,-3) to (0.2,1)  to (6.5,1)  to (v3);
	\draw [edge,->,thick] (and1) to (0,-3) to (0,1.2)  to (8.5,1.2)  to (v4);
	\end{scope}
	\end{tikzpicture}
%}
\vspace{-0.25cm}
\caption{Illustration of the reduction described in~\autoref{thm:tss_inapprox}. All arrows on the right graph represent directed edge gadgets. Thresholds are displayed inside each vertex.}
\label{fig:monCircuitReduction}
\end{figure}

\begin{enumerate}
	\item[$(i)$]  for every satisfying assignment~$A$ for~$C$ there exists a target set of size~$|A|$ for~$G$, and
	\item[$(ii)$]  for every target set~$S$ for~$G$ there exists a satisfying assignment of size~$|S|$ for~$C$.
\end{enumerate}

$(i)$ Let~$A \subseteq \{1, \ldots, n\}$ be a satisfying assignment for~$C$.
We show that the set~$S$ of vertices of~$G$ that correspond to the input nodes in~$A$ forms a target set.
%We denote with~$S$ this set of vertices.
Clearly, $|S| = |A|$.
%To see that~$S$ is indeed a target set for~$G$,
First, observe that by construction, all the~$n+1$ output vertices of~$G$  become active.
Hence, also all input vertices that are not in~$S$ become active.
Thus, all remaining vertices in~$G$ are activated since~$\thr(v) \le \deg(v)$ for all~$v \in V$.
%and removing all input nodes result in an acyclic (directed) graph.
%\todo{is this obvious or should I add more details?}
%\todo{Cristina : I'am sorry I don't understand the meaning  of of the last partstarting with "and removing...",
% I have the impression that we can just remove this part.}

$(ii)$ Let~$S \subseteq V$ be a target set for~$G$.
First, observe that we can assume that~$|S| < n$ since otherwise the satisfying assignment simply sets all input nodes to true.
Next, observe that we can assume that~$S$ is a subset of the input vertices. Indeed,
since~$G$ contains~$n+1$ copies of the circuit (excluding the input vertices), there is at least one copy without vertices in~$S$ and,
hence, the output vertex in that copy becomes active solely because of the input vertices in~$S$. %\todo{add the argument of thresholds n+1 on input nodes?}
Finally, assume by contradiction that the   set of input nodes that correspond to the vertices in~$S$ does not form a satisfying assignment.
Hence, the output node of~$C$ is evaluated to false.
However, due to the construction, this implies that the vertices corresponding to the output node are not activated,
contradicting that~$S$ is a target set for~$G$.~\end{proof}
}
\subsection{Restricted thresholds}
% \paragraph{Restricted thresholds.}
%
In this subsection, we enhance the inapproximability results to variants of \TSS with restricted threshold functions.
To this end, we use the construction desbribed in~Lemma~2 of~\cite{NNUW12} which transforms in polynomial time any instance~$I=(G = (V,E),\thr)$ of \TSS into a new instance~$I'=(G' = (V',E'),\thr')$ where $\thr'$ is the majority function such that
	\begin{enumerate}
		\item[(i)] for every target set~$S$ for~$I$ there is a target set~$S'$ for~$I'$ with~$|S'| \le |S|+1$, and
		\item[(ii)] for every target set~$S'$ for~$I'$ there is a target set~$S$ for~$I$ with~$|S| \le |S'|-1$.
	\end{enumerate}
%To this end, for majority thresholds we use the following useful lemma.
 %
%\begin{lemma}[\cite{NNUW12}, Lemma 2]\label{lem:majority equiv}
%	Let~$I=(G,\thr,k)$ be an instance of \TSS.
%	Then there is an instance~$I'=(G',\thr',k+1)$ of \TSS such that~$\thr'$ is the majority threshold  and $I$~is a yes-instance if and %only if~$I'$ is a yes-instance.
%\end{lemma}
%\todo{When we put := and when = ? I prefer to put = everywhere in Lemma 2 and 3.}
%%
%<<<<<<< .mine
%%
%%
%The construction used by \citet{NNUW12} to prove the above lemma actually ensures that $I'$ is computable in polynomial time, that for every target set~$S$ for~$I$ there is a target set~$S'$ for~$I'$ with~$|S'| \le |S|$, and for every target set~$S'$ for~$I'$ there is a target set~$S$ for~$I$ with~$|S| \le |S'|$.
%=======
%The construction used by \citet{NNUW12} to prove the above lemma actually ensures that $I'$ is computable in polynomial time, that for every target set~$S$ for~$I$ there is a target set~$S'$ for~$I'$ with~$|S'| \le |S|+1$, and for every target set~$S'$ for~$I'$ there is a target set~$S$ for~$I$ with~$|S| \le |S'|-1$.
Hence, the next corollary follows.% is an immediate consequence of this.

\begin{corollary} \label{cor:maj-hard}
 \TSS with majority thresholds is not fpt cost~$\rho$-approximable, for any computable function $\rho$, unless $\fpt=\wx{P}$.
\end{corollary}
%
%
% \subsection{Constant  thresholds}
Next, we show a similar statement for constant thresholds.

\begin{lemma}\label{lem:conmstant equiv}
	Let~$I=(G = (V,E),\thr)$ be an instance of \TSS.
	Then, we can construct in polynomial time an instance~$I'=(G' = (V',E'),\thr')$ of \TSS where
$\thr'(v) \le 2$ for all~$v\in V'$ and $G'$ is bipartite such that
	\begin{enumerate}
		%\item $I'$ can be computed in~$|I|^{O(1)}$ time, \label{item:poly time}
		%\item $\thr'(v) \le 2$ for all~$v\in V'$, \label{item:const thres}
		%\item $G'$ is bipartite, \label{item:bipartite}
		\item[(i)] for every target set~$S$ for~$I$ there is a target set~$S'$ for~$I'$ with~$|S'| = |S|$, and\label{item:solution transfer 1}
		\item[(ii)] for every target set~$S'$ for~$I'$ there is a target set~$S$ for~$I$ with~$|S| \le |S'|$. \label{item:solution transfer 2}
	\end{enumerate}
\end{lemma}
%\appendixproof{\autoref{lem:conmstant equiv}}
{
\begin{proof}
\begin{figure}[t]
\centering
\begin{tikzpicture}[scale=1, auto,swap]
%\tikzstyle{vertex}=[circle,draw,minimum size=20pt,inner sep=0pt,fill=white]

\begin{scope}
  \node[vertex, label=below:$v$] (v) at (2.5, 0) {$3$};

  \foreach \i in {1,...,4} {
      \node[vertex, label=above:$u_{\i}$] (u\i) at (\i, 1) {$1$};
      \path[edge] (v) -- (u\i);
  }
\end{scope}

\begin{scope}[xshift=6cm,yshift=0.5cm]
  \foreach \i in {1,...,4} {
      \node[vertex, label=above:$u_{\i}$] (u\i) at (\i*2-1, 2) {$1$};
      \node[vertex, label=above:$w^{\i}_{1}$] (w\i1) at (\i*2, 1) {$1$};
      \path[edge,->,thick] (u\i) -- (w\i1);

			\ifthenelse{\i > 1}{
       \node[vertex, label=above:$w^{\i}_{2}$] (w\i2) at (\i*2, -0.25) {$1$};
       \node[vertex, label=above:$\widetilde{w}^{\i}_{2}$] (tw\i2) at (\i*2 - 1, -0.25) {$2$};
       \path[edge,->,thick] (tw\i2) -- (w\i2);
       \draw [edge,->,thick] (u\i) to [bend right=25] (tw\i2);

       \pgfmathparse{int(\i-1)}  \let\r\pgfmathresult
       \path[edge,->,thick] (w\r1) -- (w\i1);
       \path[edge,->,thick] (w\r1) -- (tw\i2);
      }{}

			\ifthenelse{\i > 2}{
       \node[vertex, label=above:$w^{\i}_{3}$] (w\i3) at (\i*2, -1.50) {$1$};
       \node[vertex, label=above:$\widetilde{w}^{\i}_{3}$] (tw\i3) at (\i*2 - 1, -1.50) {$2$};
       \path[edge,->,thick] (tw\i3) -- (w\i3);
       \draw [edge,->,thick] (u\i) to [bend right=25] (tw\i3);

       \pgfmathparse{int(\i-1)}  \let\r\pgfmathresult
       \draw [edge,->,thick] (w\r2) to [bend right=25] (w\i2);
       \path[edge,->,thick] (w\r2) -- (tw\i3);
      }{}

			\ifthenelse{\i > 3}{
       \node[vertex, label=above:$w^{\i}_{4}$] (w\i4) at (\i*2, -3) {$1$};
       \node[vertex, label=above:$\widetilde{w}^{\i}_{4}$] (tw\i4) at (\i*2 - 1, -3) {$2$};
       \path[edge,->,thick] (tw\i4) -- (w\i4);
       \draw [edge,->,thick] (u\i) to [bend right=25] (tw\i4);

       \pgfmathparse{int(\i-1)}  \let\r\pgfmathresult
       \draw [edge,->,thick] (w\r3) to [bend right=25] (w\i3);
       \path[edge,->,thick] (w\r3) -- (tw\i4);
      }{}
  }

  \node[vertex, label=above:$v$] (v) at (9, -1.5) {$1$};
  \path[edge,->,thick] (w43) -- (v);

\end{scope}
\end{tikzpicture}
	\caption{Example of an activation gadget $g_v$ for a vertex~$v$ with~$\deg(v) = 4$ and~$\thr(v) = 3$. All arrows on the right graph are directed edge gadgets. Thresholds are represented inside each vertex.}
\label{fig:activation-gadget}
\end{figure}
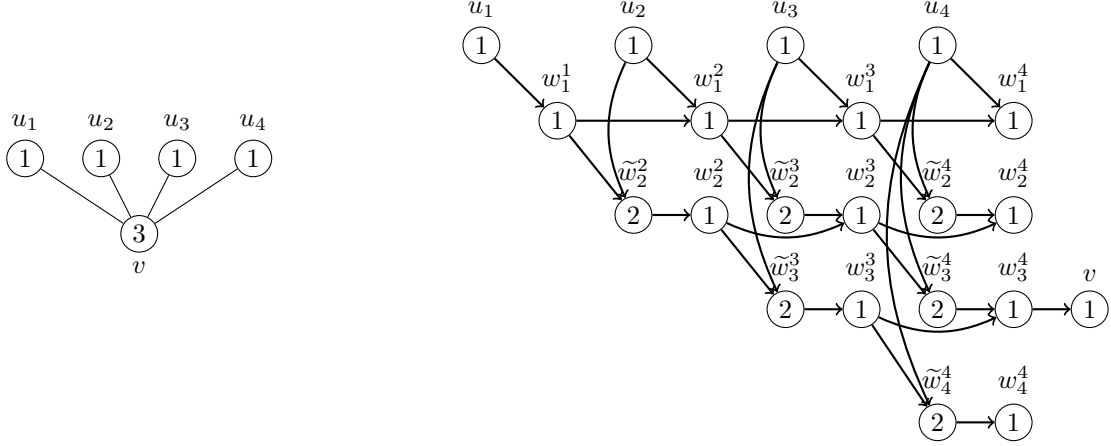
	Let~$I=(G = (V,E),\thr)$ be an instance of \TSS.
	We construct~$I' = (G' = (V',E'),\thr')$ as follows:
	Start by copying~$V$ into~$V'$, and set~$\thr'(v) = \thr(v)$ for all~$v \in V$.
% 	For each edge~$\{u,v\} \in E$ add two directed edges (using the directed edge gadget as described above) to~$G'$, one from~$u$ to~$v$ and one from~$v$ to~$u$.
	For each vertex~$v \in V'$ with~$\thr'(v) \le 2$ add for each~$u \in N_G(v)$ a directed edge gadget from~$u$ to~$v$.
	Then, for each vertex~$v \in V'$ with~$\thr'(v) > 2$ we add an \emph{activation gadget}~$g_v$ that activates~$v$ if and only if at least~$\thr(v)$ neighbors of~$v$ are active (see~\autoref{fig:activation-gadget}).
	Let~$N_G(v) = \{u_1, \ldots, u_d\}$ be the neighbors of~$v$ in~$G$.
%	Assume without loss of generality that~$\thr(v) \le d$.
%\todo{Cristina :Remove this since if we suppose that thr$\leq$ deg, this condition will be satisfied. A.N: Agree.}
% 	First, remove the directed edges from~$u_1, \ldots, u_d$ to~$v$.
	Add the vertices~$w^i_j$, $1 \le j \le i \le d$ where the vertex~$w^i_j$, $j \le i$, will become active if at least~$j$ vertices from~$\{u_1, \ldots, u_i\}$ are active.
	We inductively describe how to complete the gadget:
	First, set~$\thr(w^1_1) = 1$ and add a directed edge gadget from~$u_1$ to~$w^1_1$.
	Now, let~$i \in \{2, \ldots, d\}$.
	Then, set~$\thr(w^i_1) = 1$ and add two directed edge gadgets to~$w^i_1$, one from~$u_i$ and one from~$w^{i-1}_1$.
	For $j \in \{2, \ldots, i\}$, add a vertex~$\widetilde{w}^i_j$ with threshold two and add two directed edge gadgets to~$\widetilde{w}^i_j$, one from~$u_i$ and one from~$w^{i-1}_{j-1}$.
	Next, set~$\thr(w^i_j) = 1$ and add a directed edge gadget from~$\widetilde{w}^i_j$ to~$w^i_j$ and if~$j < i$, then add one directed edge gadget from~$w^{i-1}_j$ to~$w^i_j$.
	Observe that~$w^i_j$ becomes active if and only if~$u_i$ and~$j-1$ vertices from~$\{u_1, \ldots, u_{i-1}\}$ are active ($\widetilde{w}^i_j$ is active in this case) or if~$j$ vertices from~$\{u_1, \ldots, u_{i-1}\}$ are active ($w^{i-1}_j$ is active in this case).
	Finally, add a directed edge from~$w^d_{\thr(v)}$ to~$v$ and set~$\thr(v) = 1$.
	This completes the reduction. Clearly, $I'$ can be computed in polynomial time and $\thr'(v) \le 2$ for all~$v\in V'$.
Furthermore, observe that $G'$ is bipartite. Indeed, after removing all vertices of directed edge gadgets from~$G'$ the resulting graph contains no edge.
	Furthermore, if there is a directed edge gadget from~$u \in V'$ to~$v \in V'$, then~$u$ and~$v$ are at  distance four in $G'$.
	Hence, all cycles in~$G'$ have an even length and, thus, $G'$ is bipartite.
	%It remains to show properties~\ref{item:solution transfer 1} and~\ref{item:solution transfer 2}.

$(i)$	Let~$S \subseteq V$ be a target set for~$G$.
	We show that~$S \subseteq V'$ is also a target set for~$G'$.
	Assume by contradiction that~$S$ is not a target set for~$G'$, that is, there is at least one vertex in $G'$ that is not activated by~$S$.
	Denote by~$X$ the set of all these vertices, $X = V' \setminus \act{G',\thr'}(S)$.
	Then let~$v \in X$ be one of the earliest activated vertices in~$G$, that is, $v \in \act{G,\thr}^r(S) \setminus \act{G,\thr}^{r-1}(S)$ for some~$r \ge 1$ and~$X \cap \act{G,\thr}^{r-1}(S) = \emptyset$.
	Hence,~the set $Y=N_G(v) \cap \act{G,\thr}^{r-1}(S)$ has~$|Y|\ge \thr(v)$.
	By the choice of~$v$ it follows that all vertices in~$Y$ become active through~$S$, $Y \subseteq \act{G',\thr'}(S)$.
	If~$\thr(v) \le 2$, then~$v\in \act{G',\thr'}(S)$, a contradiction.
	If~$\thr(v) > 2$, then by the construction of the activation gadget~$g_v$, the vertex~$w^{|N_G(v)|}_{\thr(v)}$ in~$g_v$ becomes active.
	Thus~$v\in \act{G',\thr'}(S)$, a contradiction.
	This implies that~$V \subseteq \act{G',\thr'}(S)$.
	Furthermore, the set~$V$ is clearly a target set for~$G'$, that is $\act{G',\thr'}(V) = V'$, and, hence, $S$ is a target set for~$G'$.
	
%	Property \ref{item:solution transfer 2}:

$(ii)$ Let~$S'$ be a target set for~$G'$.
% 	We first show that if~$S' \setminus V \neq \emptyset$, then we can construct from~$S'$ a target set~$S$ for~$G'$ such that~$S \subseteq V$ and~$|S| \le |S'|$.
	Starting from~$S'$, we construct a target set~$S \subseteq V$ for~$G$ as follows:
	First, set~$S = S'$.
	Next, for each~$v \in S \setminus V$ remove~$v$ from~$S$ and do the following:
	If~$v$ is a vertex in a directed edge gadget from~$u$ to~$w$, then add~$u$ to~$S$.
	If~$v$ is a vertex in an activation gadget~$g_u$, then add~$u$ to~$S$.
	After exhaustively applying this procedure, we clearly have~$S \subseteq V$.
	(When a vertex~$v \in S\setminus V$ is replaced by a vertex~$w$ contained in an activation gadget~$g_u$, then~$w$ will be replaced by~$u$.)
	Next, we show that~$S$ is a target set for~$G$. Let ~$X = V \setminus \act{G,\thr}(S)$.
	Assume by contradiction that $X \neq \emptyset$.
	Then, let~$v \in X$ be one of the ``first non-activated'' vertices in~$X$, that is, $v \in \act{G',\thr'}^r(S') \setminus \act{G',\thr'}^{r-1}(S')$ for some~$r \ge 1$ and~$X \cap \act{G',\thr'}^{r-1}(S') = \emptyset$.
	Since~$S'$ is a target set for~$G'$, we have~$v \in \act{G',\thr'}(S')$ and, due to the construction of~$S$, $v \notin S'$.
	Let~$r' \ge 1$ be the integer such that~$v \in \act{G',\thr'}^{r'}(S')\setminus\act{G',\thr'}^{r'-1}(S')$.
	Let~$Y = N_G(v) \cap \act{G',\thr'}^{r'-1}(S')$.
	%Due to the construction of~$G'$ and of~$S$, we have~$|Y| \ge \thr(v)$.
	Due to the construction of $S$ and since $v$ is activated at time step $r'$ in $G'$, we know that there are at least $\thr(v)$ activated vertices among $N(v)$ in $G'$ at step $r'-1$ (using  the property of an activation gadget).	
	It follows from the choice of~$v$ that~$Y \subseteq \act{G,\thr}(S)$, and hence~$v \in \act{G,\thr}(S)$, a contradiction.
	Hence, $S'$ is a target set for~$G'$ and by construction of~$S$ it is clear that~$|S| \le |S'|$.
%	\qed
\end{proof}
}
\autoref{thm:tss_inapprox} and Lemma~\ref{lem:conmstant equiv} imply the following.

\begin{corollary} \label{cor:cst-hard}
 \TSS with thresholds at most two is not fpt cost~$\rho$-approximable even on bipartite graphs, for any computable function $\rho$, unless $\fpt=\wx{P}$.
\end{corollary}
% \begin{proof}
% 	We start with the above reduction used for the general threshold function.
% 	We now adjust the construction such that each vertex has threshold at most two.
% 	To this end, let~$v$ be a vertex with~$\thr(v) > 2$ and denote with~$u_1, \ldots, u_d$ the ``in-neighbors'' of~$v$, that is, for each~$1 \le i \le d$ there is directed edge gadget from~$u_i$ to~$v$.
% 	Observe that by construction we have~$\thr(v) = d$.
% 	Now, for all~$1 \le i \le d$ remove the directed edge gadget from~$u_i$ to~$v$.
% 	Next, add the vertices~$w_1, \ldots, w_{d-1}$ and set the threshold of these vertices to two.
% 	Then, set the threshold of~$v$ to one.
% 	Finally, add directed edges from~$u_1$ to~$w_1$, from~$u_i$ to~$w_{i-1}$, $2 \le i \le d$, and from~$w_{d-1}$ to~$v$.
%
% 	This reduction is obviously correct... \qed
% \end{proof}

\section{Parameterized inapproximability of Max  and Min $k$-Influence}\label{sec:maxminkinfluence}
%\problemopt{\maxopinfluence}{A graph $G=(V,E)$, a threshold function $\thr\colon V \to \N$, and an integer $k$.}{A subset $S \subseteq V$, $|S| \leq k$ such %that $|\sigma(S)|$ is maximum.}

%\todo[inline]{can we expand for any thresholds type? we have to control the grow of $l$...}

% We consider in this section the cardinality constraint maximization and minimization versions of \TSS. % called \maxclinfluence, \maxopinfluence, \minclinfluence, \minopinfluence.
% In these problems we search a subset of vertices of size $k$ that maximizes or minimizes the number of activated vertices.
% Their decision versions are defined as follows.
%
% \problemdec{\maxclinfluence}{A graph $G=(V,E)$, a threshold function $\thr\colon V \to \N$, and two integers $k,\ell$.}
% {Is there a subset $S \subseteq V$, $|S| = k$ such that $|\sigma[S]|\geq \ell$~?}
%
% %\problemdec{\minclinfluence}{A graph $G=(V,E)$, a threshold function $\thr\colon V \to \N$, and two integers $k,\ell$.}
% %{Is there a subset $S \subseteq V$, $|S| = k$ such that $|\sigma[S]|\leq \ell$~?}
%
%
% \maxopinfluence is defined similarly, but one ask for a solution s.t. $|\sigma(S)|\geq \ell$.
% \minclinfluence and \minopinfluence are defined also similarly, but one ask for a solution s.t. $|\sigma[S]|\leq \ell$ or $|\sigma(S)|\leq \ell$.
%
% %\maxopinfluence and \minopinfluence  are defined similarly, but one ask for a solution s.t. $|\sigma(S)]|\geq \ell$ or $|\sigma(S)]|\leq \ell$.

% We are interested in the study of the fpt cost approximation of these previous problems when the parameter is $\ell$.
We consider in this section the cardinality constraint maximization and minimization versions of \TSS.
%For the maximization version, we can show an even stronger inapproximability result when the parameter is $k+\ell$.

\begin{theorem}\label{thm:max_inapprox}
\maxclinfluence and \maxopinfluence are not fpt cost $\rho$-approximable even on bipartite graphs, for any computable function $\rho$, unless $\fpt=\wone$.
\end{theorem}
%\appendixproof{\autoref{thm:max_inapprox}}
{
\begin{proof}
 We provide a fpt gap-reduction with gap~$\rho$ from \clique to \maxclinfluence. Given an instance $I=(G=(V,E),k)$ of \clique, we construct
 an instance $I'=(G'=(V',E'), \thr, k')$ of \maxclinfluence as follows.
 The graph $G'$ consists of the incidence graph
 %\todo[inline]{MC:I think we should put directed edges between $V''$ and $E''$.
% Otherwise I could activate 2 vertices of $V''$ by selecting only one vertex in $E''$. We can also give a bigger threshold to $V''$?}
%
of $G$ with two vertex sets $V^{''}$ and $E^{''}$ corresponding to the vertices of $G$ and the edges of $G$,
 as well as a set $Z$ of $h(k)\binom{k}{2}$ new vertices $z^1_1,\ldots,z^1_{\binom{k}{2}},\ldots,z^{h(k)}_1,\ldots,z^{h(k)}_{\binom{k}{2}}$ (the function $h$ will be determined later).
 For $i=1,\ldots,\binom{k}{2}$, there is an edge directed gadget between any vertex of $E^{''}$ and vertex $z^1_i$.
%  \todo{Do we have to add a figure ?}
  Moreover, for any $i=1,\ldots,h(k)-1$, there is an edge directed gadget between  $z^i_j$  and $z^{i+1}_t$, for any $j,t\in\{1,\ldots, \binom{k}{2}\}$.
We set $k'=k$ and $g(k) = k+\binom{k}{2}+4 \binom{k}{2}^2$ (recall that there are 4 vertices per directed edge gadget). We finally set the threshold function as follows: $\thr(v) = \deg_G(v)$ for all vertices $v \in V^{''}$, $\thr(v) = 2$
   for all vertices $v \in E^{''}$ and $\thr(v) = \binom{k}{2}$ for all vertices $v \in Z$. Let $x$ be the smallest integer such that $x/\rho(x)\geq g(k)$.
   %$\frac{x}{g(x)}\geq \ell'$
    Note that such a $x$ exists and can be computed in time depending only on $k$. We choose $h$ such that $h(k)$ is an integer and  $k+ (h(k)+1)\binom{k}{2}+4h(k)\binom{k}{2}^2 \geq x$ (see~\autoref{fig:max_inapprox}).
   % \todo[inline]{Cristina: I added $h(k)$ integer and change the previous inequality. I also did some modification in the case where $G$ has no clique of size $k$. Please check.}

%\todo[inline]{The figure is quite WRONG at the moment due to the $k = 3 = \binom{3}{2}$: we can find a solution even if there is no 3-clique...}

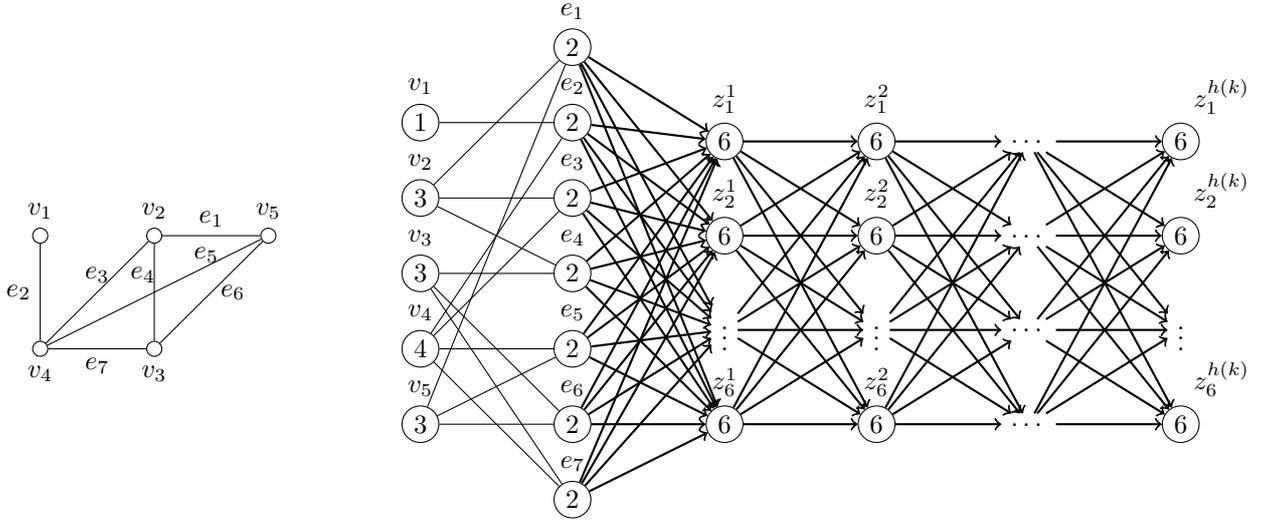
\begin{figure}[t]
\centering
\begin{tikzpicture}[scale=1, auto,swap]
%\tikzstyle{vertex}=[circle,draw,minimum size=20pt,inner sep=0pt,fill=white]

\begin{scope}[yshift=-4cm]
  \node[vertex, label=above:$v_1$] (v1) at (0, 1.5) {};
  \node[vertex, label=above:$v_2$] (v2) at (1.5, 1.5) {};
  \node[vertex, label=below:$v_3$] (v3) at (1.5, 0) {};
  \node[vertex, label=below:$v_4$] (v4) at (0, 0) {};
  \node[vertex, label=above:$v_5$] (v5) at (3, 1.5) {};
%  \path[edge] (v1) -- node[above] {$e_1$} (v2);
  \path[edge] (v2) -- node[above] {$e_1$} (v5);
  \path[edge] (v4) -- node[left] {$e_2$} (v1);
  \path[edge] (v2) -- node[above] {$e_3$} (v4);
  \path[edge] (v2) -- node[above,xshift=-0.15cm] {$e_4$} (v3);
  \path[edge] (v4) -- node[right,xshift=0.4cm,,yshift=0.5cm] {$e_5$} (v5);
  \path[edge] (v3) -- node[right] {$e_6$} (v5);
  \path[edge] (v3) -- node[below] {$e_7$} (v4);
\end{scope}

\begin{scope}[xshift=4cm]

  \foreach \i/\t in {1/1,2/3,3/3,4/4,5/3} {
      \node[vertex, label=above:$v_\i$] (v\i) at (1,-\i) {$\t$};
  }

  \foreach \i in {1,...,7} {
    \node[vertex, label=above:$e_\i$] (e\i) at (3,-\i+1) {$2$};
  }

  \path[edge] (v2) -- (e1) -- (v5);
  \path[edge] (v1) -- (e2) -- (v4);
  \path[edge] (v2) -- (e3) -- (v4);
  \path[edge] (v2) -- (e4) -- (v3);
  \path[edge] (v5) -- (e5) -- (v4);
  \path[edge] (v5) -- (e6) -- (v3);
  \path[edge] (v4) -- (e7) -- (v3);

  \foreach \j in {1,...,4} {
    \foreach \i in {1,...,4} {
			\ifthenelse{\j = 3}{
        \node (z\i\j) at (2*\j+3,-\i*1.25) {\ldots};
      }
      {
			  \ifthenelse{\j = 4}{
			    \ifthenelse{\i = 3}{
            \node (z\i\j) at (2*\j+3,-\i*1.25) {\vdots};
          }{
			      \ifthenelse{\i = 4}{
              \node[vertex, label=80:$z_6^{h(k)}$] (z\i\j) at (2*\j+3,-\i*1.25) {$6$};
            }{
              \node[vertex, label=80:$z_\i^{h(k)}$] (z\i\j) at (2*\j+3,-\i*1.25) {$6$};
            }
          }
        }
        {
			    \ifthenelse{\i = 3}{
            \node (z\i\j) at (2*\j+3,-\i*1.25) {\vdots};
          }{
			      \ifthenelse{\i = 4}{
              \node[vertex, label=above:$z_6^\j$] (z\i\j) at (2*\j+3,-\i*1.25) {$6$};
            }{
              \node[vertex, label=above:$z_\i^\j$] (z\i\j) at (2*\j+3,-\i*1.25) {$6$};
            }
          }
        }
      }
    }
  }

  \foreach \k in {1,...,3} {
    \pgfmathparse{int(\k+1)}  \let\r\pgfmathresult
    \foreach \i in {1,...,4} {
      \foreach \j in {1,...,4} {
       \path[edge,->,thick] (z\i\k) -- (z\j\r);
      }
    }
  }

  \foreach \i in {1,...,7} {
    \foreach \j in {1,...,4} {
       \path[edge,->,thick] (e\i) -- (z\j1);
    }
  }

\end{scope}
\end{tikzpicture}
\caption{Illustration of the reduction from \clique described in~\autoref{thm:max_inapprox} with~$k=4$. All arrows on the right graph are directed edge gadgets. Thresholds are represented inside each vertex.}
\label{fig:max_inapprox}
\end{figure}

If $G$ contains a clique of size $k$, then, by activating the same $k$ vertices in $V^{''}$,
 we activate in the next step exactly $\binom{k}{2}$ vertices in $E^{''}$ (the vertices correspond to the edges of the clique).
 During the next steps, all the vertices of $Z$ are activated. Overall, $opt(I') \geq k + (h(k)+1)\binom{k}{2} + 4 h(k)\binom{k}{2}^2 \geq x$.
  Moreover, $g(k)\leq x/\rho(x) \leq opt(I')/\rho(opt(I'))$,
  %$\ell'\leq \nicefrac{x}{g(x)} \leq \nicefrac{opt(I')}{g(opt(I'))}$
   since the function $t/\rho(t)$
%  $\nicefrac{t}{g(t)}$
   is nondecreasing.
  Otherwise, suppose that there is no clique of size $k$ in $G$. Without loss of generality, we may assume that~$k \geq 4$ and thus~$k < \binom{k}{2}$. Remark that we may also assume that no vertex from $Z$ is in
  an optimal solution since it activates only $4\binom{k}{2}$ more vertices from gadgets and since the threshold of vertices of $Z$ is $\binom{k}{2}$, no more vertex from $Z$ can be activated. From the same reason, no vertex from the edge directed gadgets is in an optimal solution. Thus, only vertices from  $V^{''} \cup E^{''}$ are
  in the optimal solution. %By the definition of the incidence graph and the thresholds,
 % \todo{as morgan said, not really, since activate $k$ in $E''$ can activate up to $2k$ in $V''$ and then,...}
%  When $t$ vertices are activated in $E^{''}$ and $k-t$ in $V^{''}$, then in the first step some new vertices are activated in $V^{''}$ but there is no way  to activate  $\binom{k}{2}$ vertices in   $E^{''}$.
  Suppose that~$t \geq 0$ vertices~$S_1 = \{e_1',\ldots,e_t'\}$ are activated in $E^{''}$ and~$k-t$ vertices~$S_2 = \{v_1',\ldots,v_{k-t}'\}$ in $V^{''}$.
Then there is no way  to activate~$\binom{k}{2}$ vertices in~$E^{''}$. Assume by contradiction that this is possible.
%there exists a smallest integer~$r > 0$ such that~$|\act{G',\thr}^{r}(S) \cap E''| \geq \binom{k}{2}$  where~$S = S_1 \cup S_2$. Since~$S \subseteq V'' \cup E''$ and every vertex of~$V'' \cup E''$ has unanimity threshold it follows that~$r$ must be less or equal to~$1$.
If~$t=0$ then~$|\act{G',\thr}^{1}(S) \cap E''| < \binom{k}{2}$ since there is no clique of size~$k$ in~$G$, a contradiction. If~$t=k$ then ~$|\act{G',\thr}^{1}(S) \cap E''| \leq k < \binom{k}{2}$ by assumption, a contradiction. Thus~$1\leq t<k$ and then~$|\act{G',\thr}^{1}(S) \cap E''| \leq \binom{k-t}{2} + t <  \binom{k}{2}$, a contradiction.
%Since the vertices of~$V''$ have unanimity thresholds it follows that every vertex~$v \in V''$ activated by propagation is such that~$N_{G'}(v) \subseteq S_1$
Thus $opt(I') < k + \binom{k}{2} + 4 \binom{k}{2}^2=g(k)$.
The result follows from Lemma~\ref{lem:fpt-gap} together with the \wone-hardness of~\clique~\cite{DF99}.
%    Hence, a $g(\ell')$-approximation algorithm for \maxclinfluence applied to $I'$ could distinguish in fpt-time between
%   \textit{yes}-instances and \textit{no}-instances for \clique, implying $\fpt=\wone$ since this last problem is \wone-hard~\cite{DF99}.

   The same reduction works for \maxopinfluence except that $g(k)= \binom{k}{2}+4 \binom{k}{2}^2$ and that $h$ is defined such that $(h(k)+1)\binom{k}{2}+4h(k)\binom{k}{2}^2 \geq x$.
%\qed
\end{proof}
}
We remark that the proof of \autoref{thm:max_inapprox} shows a stronger result:
Unless $\fpt=\wone$, there is no fpt cost $\rho$-approximation for \maxclinfluence and \maxopinfluence, for any computable function $\rho$, even if the running time is of the form~$f(k,\ell) \cdot n^{O(1)}$. Here $\ell$ is the cost-parameter passed as an argument to the algorithm, that is, $\ell$ indicates the number of activated vertices.
% Indeed, the proof shows the hardness even in the case where~$k$ solely depends on $\ell$.\todo{is it clear?! MC: is it better? / Do you think people understand what is $\ell$? I am not sure...according to the def of a fpt cost approx algo I think it is understandable.}
%
%
%
% \todo[inline]{Let as open question the result for other thresholds and explain why the current 2 lemmas does not hold as they are due to the increase of $l$}
%
%As the reductions behind the Corollaries~\ref{cor:maj-hard} and~\ref{cor:cst-hard}
%are not parameterized reductions with respect to the number~$\ell$ of activated vertices, we cannot apply these lemmas to prove the same  cost inapproximability results for \maxclinfluence or \maxopinfluence with majority thresholds and thresholds bounded by 2.

As the reductions behind  Corollaries~\ref{cor:maj-hard} and~\ref{cor:cst-hard}
are not fpt gap-reductions, we cannot use them to prove the same  cost inapproximability results for \maxclinfluence or \maxopinfluence with majority thresholds and thresholds at most two.
\paragraph{Minimization variants.}
% Consider in the following the minimization versions.

In this following, we study minimization versions of the problem. One practical application could be in security of social networks, for example, finding the $k$ most trustworthy individuals such that the leak of critical information is minimized.

In contrast to the maximization versions which remain NP-hard for unanimity thresholds~\cite{BCNS12}, we can show that the minimization variants are polynomial-time solvable
for unanimity thresholds.

\begin{proposition}\label{prop:minUnanimityPoly}
\minopinfluence and \minclinfluence are solvable in polynomial time for unanimity thresholds.
\end{proposition}
%\appendixproof{\autoref{prop:minUnanimityPoly}}
{
\begin{proof} Consider any instance of \minopinfluence. We have $opt=0$ or 1 and $opt=1$ if and only if $k=n-1$.  In order to see this,
we distinguish two cases depending on the connectivity of $G$. If $G$ is connected,  let $T$ be a spanning tree of $G$.
An optimal solution $S$ is constructed by adding iteratively
 a vertex $v$ %\todo{a leaf $v$?}
 in $S$ such that $T-\{v\}$ remains a spanning tree of $G-\{v\}$.
 If $k<n-1$, then any vertex in $V\setminus S$ has a neighbor in $V\setminus S$ and so $opt=0$.
 If $G$ is not connected and $k<n-1$, a solution of value 0 always exists and consists
 of all the vertices of a connected component or all the vertices of a connected component except  at least two connected vertices.
 Clearly, \minclinfluence is also solvable in polynomial time.
 %\todo[inline]{I would like to assume that $deg(v) >0$ for any vertex and 1 $\leq \thr(v) \leq deg(v)$. DO you agree ? A.N: Agree.}
% \todo{F: Make the small corrections to be clearer; C: Done}
%\qed
\end{proof}
}

The next result shows that \minclinfluence and \minopinfluence are also computationally hard even for thresholds bounded by two.
To this end, we consider the decision version of \minclinfluence (resp. \minopinfluence) denoted by \decminclinfluence (resp. \decminopinfluence) and defined as follows: Given a graph $G=(V,E)$, a threshold function $\thr\colon V \to \N$, and integers $k$ and~$\ell$, determine whether there is a  subset $S \subseteq V$, $|S| = k$ such that $|\act{G,\thr}(S)| \leq \ell$ (resp. $|\act{G,\thr}(S) \setminus S| \leq \ell$).

\begin{theorem}\label{thm:min_whard}
\decminclinfluence is \wone-hard w.r.t. parameter $(k,\ell)$ even for threshold bounded by two and bipartite graphs.
\decminopinfluence is $\np$-hard even for threshold bounded by two, bipartite graphs and $\ell=0$. %\todo{Not in XP for param $l$?}
%\minclinfluence is \wone-hard w.r.t. parameter $k+ \ell$ even for threshold bounded by 2 and  bipartite graphs.
% \minopinfluence for threshold bounded by 2 and  bipartite graphs is $\np$-hard even for $\ell=0$. %\todo{Not in XP for param $l$?}
\end{theorem}
%\appendixproof{\autoref{thm:min_whard}}
{
\begin{proof}
We provide a parameterized reduction from the \wone-hard problem \stable to \decminclinfluence.
Given an instance $(G=(V,E), k)$
of \stable, we construct an instance $(G'=(V',E'),k',\ell')$  of \decminclinfluence by considering the incidence graph.
Let $G'$ be the bipartite graph with two vertex sets $V$ and $E$ and for each edge $e=uv \in E$, we add  $ue, ve \in E'$.
We define $\thr(u) =1, \forall u \in V$ and $\thr(e) = 2, \forall e \in E$, $k'=k$ and $\ell'=k$. Clearly $G$ contains an independent set of size at least $k$
if and only if $G'$ contains a set of vertices $S$ of size $k$ such that $|\act{G,\thr}(S)|= k$.

The  previous construction also prove that $G$ contains an independent set of size at least $k$
if and only if $G'$ contains a set of vertices $S$ of size $k$ such that  $|\act{G,\thr}(S) \setminus S|= 0$,
thus proving the $\np$-hardness of \decminopinfluence even for $\ell=0$.
%\qed
\end{proof}
}

We remark that the previous theorem rules out the possibility of any fixed-parameter algorithm with parameter $\ell$ for \decminopinfluence assuming $\p \neq \np$.
Moreover, due to its $\np$-hardness when $\ell=0$, \minopinfluence is not at all fpt cost approximable, unless $\p = \np$.

%In the following, we study the fpt cost approximation of \minclinfluence depending on parameter $\ell$. %\todo{say something for \minopinfluence?}
In the following, we provide a final result regarding fpt cost approximation of \minclinfluence.

\begin{theorem}\label{thm:min_inapprox}
%There is no fpt cost $g(\ell)$-approximation algorithm parameterized by $\ell$  for \minclinfluence even for threshold bounded by 2  and  bipartite graphs, for any computable function $g$, unless $\fpt=\wone$.
\minclinfluence with thresholds at most two is not fpt cost $\rho$-approximable even on bipartite graphs, for any computable function $\rho$, unless $\fpt=\wone$.
\end{theorem}
%\appendixproof{\autoref{thm:min_inapprox}}
{
\begin{proof}

%\appendixproof{Figure for \autoref{thm:min_inapprox}}
{
\begin{figure}[t]
\centering
\begin{tikzpicture}[scale=1, auto,swap]
%\tikzstyle{vertex}=[circle,draw,minimum size=20pt,inner sep=0pt,fill=white]

\begin{scope}[yshift=-3cm]
  \node[vertex, label=above:$v_1$] (v1) at (0, 1) {};
  \node[vertex, label=above:$v_2$] (v2) at (1, 1) {};
  \node[vertex, label=below:$v_3$] (v3) at (1, 0) {};
  \node[vertex, label=below:$v_4$] (v4) at (0, 0) {};
  \path[edge] (v1) -- node[above] {$e_1$} (v2);
  \path[edge] (v2) -- node[right] {$e_2$} (v3);
  \path[edge] (v3) -- node[below] {$e_3$} (v4);
  \path[edge] (v4) -- node[left] {$e_4$} (v1);
  \path[edge] (v2) -- node[above] {$e_5$} (v4);
\end{scope}

%\begin{scope}[xshift=10cm,rotate=-90]
\begin{scope}[xshift=4cm]
  \foreach \i in {1,...,4} {
      \node[vertex, label=above:$v_\i$] (v\i) at (0,-\i-0.5) {$1$};
  }

  \foreach \i in {1,...,5} {
    \node[vertex, label=above:$e_\i$] (e\i) at (2,-\i) {$2$};
  }

  \path[edge] (v1) -- (e1) -- (v2);
  \path[edge] (v2) -- (e2) -- (v3);
  \path[edge] (v3) -- (e3) -- (v4);
  \path[edge] (v4) -- (e4) -- (v1);
  \path[edge] (v2) -- (e5) -- (v4);

  \node[vertex, label=right:$z_1$] (z1) at (4,0) {$1$};
  \node[vertex, label=right:$z_2$] (z2) at (4,-1) {$1$};
  \node (vdots) at (4,-3) {$\ldots$};
  \node[vertex, label=right:$z_{h(k)}$] (zg) at (4,-6) {$1$};

  \foreach \i in {1,...,5} {
    \draw (e\i) -- (z1);
    \draw (e\i) -- (z2);
    \draw (e\i) -- (vdots);
    \draw (e\i) -- (zg);
  }

\end{scope}
\end{tikzpicture}
\caption{Illustration of the reduction from \stable as described in~\autoref{thm:min_inapprox}. Thresholds are represented inside each vertex.}
\label{fig:min_inapprox}
\end{figure}
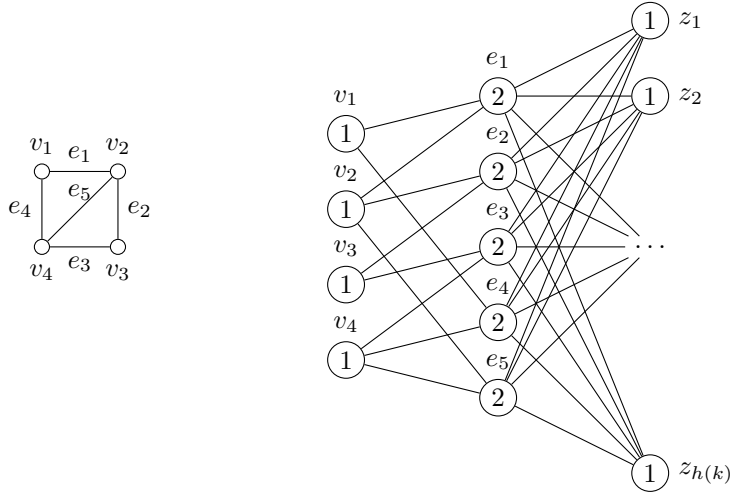
}

We provide a fpt gap-reduction with gap~$\rho$ from \stable to \minclinfluence (see also \autoref{fig:min_inapprox}).
Given an instance $(G=(V,E), k)$
of \stable, we construct an instance $(G'=(V',E'),k')$  of \minclinfluence by considering the incidence graph, that is $G'$ is a bipartite graph with two vertex sets $V$ and $E$ and for each edge $e=uv \in E$, there is $ue, ve \in E'$.
We define $\thr(u) =1, \forall u \in V$ and $\thr(e) = 2, \forall e \in E$. We choose the function $h$ such that $h(k)$ is an integer and  $k+h(k)+1 \geq k \rho(k)$. %\todo{Cristina: The equality becomes inequality and add h(k) integer.}
Then, we add  $h(k)$   additional
vertices $F$   of threshold 1 in $G'$ and a complete bipartite graph between $E$ and $F$. Define $k'=k$, $g(k)=k+h(k)+1$.
If $G$ contains an independent set of size at least $k$ then, by activating the same $k$ vertices in $G'$,
 we obtain a solution that activates no more vertex in $G'$ and thus  $opt(I')=k\leq \frac{g(k)}{\rho(k)}=\frac{g(k)}{\rho(opt(I'))}$.
% $opt(I')=k=\nicefrac{\ell'}{g(k)}=\nicefrac{\ell'}{g(opt(I'))}$

If there is no independent set of size $k$, if one activate only two vertices from $F$,  it will activate the whole vertex set $E$ on the next step, and then the whole graph. Moreover, activating a vertex from the vertex set $E$ will also activate the whole set $F$ on the next step, and then the whole graph. Finally, activating $k$ vertices of $V$ will activate at least one vertex of $E$ since there is no independent set of size $k$. Note that activating $k-1$ vertices from $V$ and 1 from $F$ will result not be better since vertices of $F$ are connected to all vertices of $E$. Therefore, $opt(I')\geq k+h(k)+1 =g(k)$.

The result follows from Lemma~\ref{lem:fpt-gap} together with the \wone-hardness of~\stable~\cite{DF99}.
%Hence, a $g(\ell')$-approximation algorithm for \minclinfluence applied to $I'$ could distinguish in fpt-time between
%   \textit{yes}-instances and \textit{no}-instances for \stable, implying $\fpt=\wone$ since this last problem is \wone-hard~\cite{DF99}.
%\qed
\end{proof}
}

%\todo[inline]{maybe the lemma holds for majority since it's min.
%
%A: No, the number of activated vertices in the new instance depends on the degree, which is to high in the construction.}

%\section{Dual TSS}\label{sec:dualtss}
%  \problemparam{\dualTSS (Dual TSS)}{An undirected graph~$G=(V,E)$, a threshold function $\thr\colon V\rightarrow\N\cup\{0\}$ and an integer~$k$.}{$k$}{Is there a
%  target set~$S\subseteq V$ for~$G$ that contains at most~$n-k$ vertices?}
%
%
%  Using the fact that TSS  with unanimity thresholds is exactly Vertex Cover, we can show that \begin{theorem}
%   \dualTSS with unanimity thresholds is W[1]-hard w.r.t. parameter $k$.
%\end{theorem}
%
%
%       We are trying to study the parametrized complexity of Dual  TSS in the case of    thresholds
%       bounded by a constant and if hard then study the parameterized approximability.
%
%
% \vspace{-0.5cm}

\section{Conclusion} \label{sec:conclusions}

%In this paper, we enforced the fact that MinTSS is computationally very hard. Indeed, even in fpt-time, there is no $g(k)$-approximation for this problem, even with restrictive thresholds.

Despite the variety of our intractability results, some questions remains open.
Are \maxclinfluence and \maxopinfluence fpt cost approximable for constant or majority thresholds?
We believe that these problems remain hard, but the classical gadgets used to simulate these thresholds changes does not work for this type of approximation.
Similarly, is \minclinfluence fpt cost approximable for majority thresholds?

Finally, the dual problem of  \TSS (i.e. find a target set of size at most $|V|-k$) seems unexplored. Using the
 fact that \TSS with unanimity thresholds is exactly Vertex Cover, we know that the dual problem is
  therefore $\wone$-hard, even with unanimity thresholds. But it is still the case for constant or majority thresholds? Moreover,
   is the dual of \TSS fpt cost approximable?

{%\footnotesize
	\bibliography{bib}

\begin{thebibliography}{10}

\bibitem{aazami2009}
Ashkan Aazami and Kael Stilp.
\newblock Approximation algorithms and hardness for domination with
  propagation.
\newblock {\em SIAM J Discrete Math}, 23(3):1382--1399, 2009.

\bibitem{AL96}
Sanjeev Arora and Carsten Lund.
\newblock Hardness of approximations.
\newblock In {\em Approximation algorithms for \textit{NP}-hard problems},
  pages 399--446. PWS Publishing Company, 1996.

\bibitem{BCNS12}
Cristina Bazgan, Morgan Chopin, Andr{\'{e}} Nichterlein, and Florian Sikora.
\newblock Parameterized approximability of maximizing the spread of influence
  in networks.
\newblock {\em J. Discrete Algorithms}, 27:54--65, 2014.

\bibitem{benzwi2011}
Oren Ben-Zwi, Danny Hermelin, Daniel Lokshtanov, and Ilan Newman.
\newblock Treewidth governs the complexity of target set selection.
\newblock {\em Discrete Optim}, 8(1):87--96, 2011.

\bibitem{CH2010}
Liming Cai and Xiuzhen Huang.
\newblock Fixed-parameter approximation: Conceptual framework and
  approximability results.
\newblock {\em Algorithmica}, 57(2):398--412, 2010.

\bibitem{chang2009}
Ching-Lueh Chang and Yuh-Dauh Lyuu.
\newblock Spreading messages.
\newblock {\em Theor Comput Sci}, 410(27--29):2714--2724, 2009.

\bibitem{chen2009}
Ning Chen.
\newblock On the approximability of influence in social networks.
\newblock {\em SIAM J Discrete Math}, 23(3):1400--1415, 2009.

\bibitem{chen2006}
Yijia Chen, Martin Grohe, and Magdalena Gr{\"u}ber.
\newblock On parameterized approximability.
\newblock In Hans~L. Bodlaender and Michael~A. Langston, editors, {\em Proc. of
  IWPEC '06}, volume 4169 of {\em LNCS}, pages 109--120. Springer, 2006.

\bibitem{CNNW12}
Morgan Chopin, Andr{\'e} Nichterlein, Rolf Niedermeier, and Mathias Weller.
\newblock Constant thresholds can make target set selection tractable.
\newblock In Guy Even and Dror Rawitz, editors, {\em Proc. of MedAlg '12},
  volume 7659 of {\em LNCS}, pages 120--133. Springer, 2012.

\bibitem{Cicalese2013}
Ferdinando Cicalese, Gennaro Cordasco, Luisa Gargano, Martin Milanic, and Ugo
  Vaccaro.
\newblock Latency-bounded target set selection in social networks.
\newblock {\em Theor. Comput. Sci.}, 535:1--15, 2014.

\bibitem{DS02}
Irit Dinur and Shmuel Safra.
\newblock The importance of being biased.
\newblock In John~H. Reif, editor, {\em Proceedings on 34th Annual ACM
  Symposium on Theory of Computing (STOC'02)}, pages 33--42. ACM, 2002.

\bibitem{DF99}
Rodney~G. Downey and Michael~R. Fellows.
\newblock {\em Fundamentals of Parameterized Complexity}.
\newblock Springer, 2013.

\bibitem{downey2006}
Rodney~G. Downey, Michael~R. Fellows, and Catherine McCartin.
\newblock Parameterized approximation algorithms.
\newblock In Hans~L. Bodlaender and Michael~A. Langston, editors, {\em Proc. of
  IWPEC '06}, volume 4169 of {\em LNCS}, pages 121--129. Springer, 2006.

\bibitem{dreyer2009}
Paul~A. Dreyer and Fred~S. Roberts.
\newblock Irreversible $k$-threshold processes: Graph-theoretical threshold
  models of the spread of disease and of opinion.
\newblock {\em Discrete Appl Math}, 157(7):1615 -- 1627, 2009.

\bibitem{FG06}
J{\"o}rg Flum and Martin Grohe.
\newblock {\em Parameterized Complexity Theory}.
\newblock Springer, 2006.

\bibitem{kempe2003}
David Kempe, Jon Kleinberg, and {\'{E}}va Tardos.
\newblock Maximizing the spread of influence through a social network.
\newblock In Lise Getoor, Ted~E. Senator, Pedro Domingos, and Christos
  Faloutsos, editors, {\em Proc. of the 9th SIGKDD International Conference on
  Knowledge Discovery and Data Mining (KDD '03)}, pages 137--146. ACM, 2003.

\bibitem{marx06}
Daniel Marx.
\newblock Parameterized complexity and approximation algorithms.
\newblock {\em Comput J}, 51(1):60--78, 2008.

\bibitem{Mar13}
D{\'a}niel Marx.
\newblock Completely inapproximable monotone and antimonotone parameterized
  problems.
\newblock {\em J. Comput. Syst. Sci.}, 79(1):144--151, 2013.

\bibitem{NNUW12}
Andr\'e Nichterlein, Rolf Niedermeier, Johannes Uhlmann, and Mathias Weller.
\newblock On tractable cases of target set selection.
\newblock {\em Soc Network Anal Mining}, 3(2):233--256, 2013.

\bibitem{Nie06}
Rolf Niedermeier.
\newblock {\em Invitation to Fixed-Parameter Algorithms}.
\newblock Oxford University Press, 2006.

\bibitem{peleg02}
David Peleg.
\newblock Local majorities, coalitions and monopolies in graphs: a review.
\newblock {\em Theor. Comput. Sci.}, 282(2):231--257, 2002.

\bibitem{reddy2011}
T.~V.~Thirumala Reddy and C.~Pandu Rangan.
\newblock Variants of spreading messages.
\newblock {\em J Graph Algorithms Appl}, 15(5):683--699, 2011.

\end{thebibliography}
	\bibliographystyle{plain}
}
% \bibliographystyle{abbrv}
%\bibliography{main}

%\nocite{*}

%no journal
\appendix
%\newpage
%\section{Appendix}
%\appendixProofText

\end{document}